\documentclass[11pt,a4paper]{article}
\pdfoutput=1
\usepackage{jheppub}
\usepackage[utf8]{inputenc}
\usepackage{amssymb, amsmath}
\usepackage[normalem]{ulem}
\usepackage{hyperref}
\usepackage{graphicx}
\usepackage{subfigure}

\newcommand{\bea}{\begin{eqnarray}}
\newcommand{\eea}{\end{eqnarray}}
\renewcommand{\d}{{\mathrm{d}}}

\def\be{\begin{equation}}
\def\ee{\end{equation}}

\def\doi{http://doi.org}

\def\d{\mathrm{d}}

\begin{document}
	
	\title{Stochastic eternal inflation is in the swampland} 	

	\author{Suddhasattwa Brahma$^{a,*}$ and}
	\emailAdd{$^*$suddhasattwa.brahma@gmail.com}
	\affiliation{$^{a}$Asia Pacific Center for Theoretical Physics, Pohang 37673, South Korea}
	
	\author{Sarah Shandera$^{b,\dagger}$}
	\emailAdd{$^\dagger$ses47@psu.edu}
	\affiliation{$^{b}$Institute for Gravitation and the Cosmos, The Pennsylvania State University, University Park, PA 16802, USA}
	
	\date{August 09, 2019}

\abstract{We demonstrate that there is no controlled description of stochastic eternal inflation consistent with the refined swampland de Sitter conjecture.}

\maketitle

\section{Introduction}

The refined de Sitter (dS) conjecture of \cite{Garg:2018reu,Ooguri:2018wrx} proposes a constraint on the nature of quasi-dS solutions consistent with a full theory of quantum gravity. For our purposes, the conjecture states that any quasi-dS space sourced by potential energy density $V$ that is a function of a single scalar degree of freedom, where
\begin{equation}
|V^{\prime}|<\frac{c}{M_p}V
\label{eq;conjecture1}
\end{equation}
for $c>0$ and order one, must satisfy
\begin{equation}
V^{\prime\prime}\leq\frac{-\tilde{c}}{M_p^2}V\,.
\label{eq;conjecture2}
\end{equation}
Here $\tilde{c}$ is also a positive constant of order 1 and $M_p$ is the reduced Planck mass. Ooguri et al \cite{Ooguri:2018wrx} demonstrated that the compatibility of semi-classical spacetimes undergoing accelerated expansion, Eq.(\ref{eq;conjecture1}), the covariant entropy bound \cite{Bousso:1999xy,Banks:2000fe}, and the distance conjecture \cite{Ooguri:2006in} requires the condition in Eq.~(\ref{eq;conjecture2}) above. Although the original de Sitter (dS) swampland conjecture was motivated by the difficulty of finding meta-stable dS solutions in string theory, the argument of \cite{Ooguri:2018wrx} leads to the same conclusion starting from more general principles of quantum gravity, namely the distance conjecture together with a semi-classical notion of entropy for quasi-dS spacetimes.

The instability in the potential required by Eq.(\ref{eq;conjecture2}) of the refined dS conjecture restricts the amount of inflation that any one observer can see. One might wonder if an inflationary era could be stretched beyond this limit by a period of stochastic eternal inflation \cite{Vilenkin:1983xq,Linde:1986fc,Linde:1986PLB}, where quantum fluctuations are strong enough to balance or overcome the classical field motion along the potential and maintain the field in a region of the potential that supports inflation. If the quantum fluctuations are strong enough, they may in principle also change the global nature of the spacetime by generating sufficient inflating volumes that the global volume is dominated by those regions even as inflation ends in some Hubble patches. 

However, from at least one perspective it would be surprising if the stochastic eternal inflation phenomena is consistent with the dS conjectures: the quantum fluctuations in the effective theory would be allowing a violation of constraints imposed by the UV quantum gravity theory. The extra inflation cannot be entirely for free. Even if the competition between classical motion and quantum fluctuations were to pin the field's expectation value at a point, high energy modes of the fluctuating field would be continually red-shifted into the low energy effective theory. It was previously shown by \cite{Matsui:2018bsy, Dimopoulos:2018upl} that a stronger, earlier version of the dS conjecture \cite{Obied:2018sgi} restricted some forms of stochastic eternal inflation. Here we will show that stochastic eternal inflation, when it is under perturbative control, is entirely incompatible with even the refined dS conjecture. We demonstrate this by enforcing a consistency constraint on the eternal inflation regime that was overlooked by \cite{Kinney:2018kew}.

\section{Requirements for a perturbative description of stochastic eternal inflation:}

In stochastic eternal inflation, the quantum motion of the field sourcing inflation is large enough to compete with the classical evolution of the field. The Hubble parameter, $H$, during inflation is given by $3H^2M_p^2=V(\phi)$, and $H\ll M_p$ is required to keep the scenario in the regime of semi-classical gravity. The equation of motion for the scalar field, in the limit where acceleration is small, is $3H\dot{\phi}=-c_sV^{\prime}(\phi)$, where dots indicate time derivatives, $V^{\prime}=\frac{\d V}{\d\phi}$, and we allow sound speed $c_s$ different from 1 for the inflaton. In general, the field evolution from quantum fluctuations dominate over the classical motion when
\begin{align}
\frac{\langle \delta \phi\rangle_{\rm Q}}{\delta\phi_{\rm Cl}}=\frac{H^2}{2\pi\dot\phi}>1\,,
\label{eq:dQdcl}
\end{align}
where we have used $\langle \delta \phi\rangle_{\rm Q}$ for a massless, free scalar field. In single-field inflation, perturbations of the metric are sourced by the perturbations in the scalar field through the Einstein equation. There are different useful choices of gauge to describe the fluctuations of the single physical (scalar) degree of freedom. When computing observables after inflation it is convenient to use comoving gauge where $\delta\phi=0$ and the physical fluctuation is carried in the curvature perturbations, $\zeta$. In this case the metric, including scalar perturbations, is $ds^2=-\d t^2+a^2(t)(1+2\zeta(\vec{x},t))\d\vec{x}^2$.  Here $a$ is the homogeneous scale factor, so $H=\dot{a}/a$. Then the quadratic action for the perturbations is
\begin{equation}
S_2=M_p^2\int \d t\,\d^3 x\left(a^3\frac{\epsilon}{c_s^2}\dot{\zeta}^2-a\epsilon(\partial\zeta)^2\right)\,.
\end{equation}
We have allowed for the possibility of a non-trivial sound speed for the inflaton, as would be generated by the presence of higher derivative terms in the inflaton action. Such terms can be generated if the inflaton field, whose action should be nearly shift-symmetric to support inflation, is coupled to heavy fields (not subject to such a restriction) that can been integrated out. The size of the observable inhomogeneities after inflation is set by the dimensionless power spectrum, $\mathcal{P}_{\zeta}$, where the full amplitude of fluctuations in mode $k$ is $P(k)=\frac{2\pi^2\mathcal{P}_{\zeta}}{k^3}$. The dominant contribution to $\mathcal{P}_{\zeta}$ is computed from the quadratic action above.

The curvature perturbation $\zeta$ is related to the scalar fluctuation $\delta\phi$ in the spatially flat gauge by $\zeta=-\frac{H}{\dot{\phi}}\delta\phi$. The condition in Eq.(\ref{eq:dQdcl}) then says, in comoving gauge, that stochastic eternal inflation can occur when the dimensionless amplitude of curvature perturbations is at least of order one:
\begin{align}
\mathcal{P}_{\zeta}=\left(\frac{H^2}{2\pi\dot{\phi}}\right)^2=\frac{H^2}{8\pi^2M_p^2\epsilon_V c_s}>1\,,
\end{align}
where we have introduced the slow-roll parameter
\begin{equation}
\epsilon_V=\frac{M_p^2}{2}\left(\frac{V^{\prime}}{V}\right)^2\,.
\end{equation}
Note that the condition for accelerated expansion is $\epsilon_V<1$, which is the same as Eq.(\ref{eq;conjecture1}) with $c=\sqrt{2}$. 
Although the curvature perturbations are large, the regime of stochastic eternal inflation can still be described perturbatively as long as corrections to the two-point function from higher order terms in the action are small \cite{Creminelli:2008es,Leblond:2008gg}. The full computation of the correlation functions is somewhat involved, but well-known. However, for single-field, slow-roll inflation the approximate conditions for perturbativity can be read off from the action by requiring all terms with $n$ powers of $\zeta$ to be smaller than the quadratic term ($S_n/S_2<1$) when substituting $\partial/\partial t\rightarrow H$, $\partial\rightarrow aH$ and $\zeta\rightarrow\mathcal{P}_{\zeta}^{1/2}$. This procedure works since the correlation functions in single-field are determined at horizon crossing \cite{Shandera:2008ai}.

In the simplest case where all slow-roll parameters scale like powers of $\epsilon$, slow-roll inflation can be treated perturbatively as long as
\begin{equation}
\mathcal{P}_{\zeta}<\frac{1}{\epsilon_V^2}\,.
\label{eq:perturbeps}
\end{equation}
This allows at least some cases of stochastic eternal inflation, with $\mathcal{P}_{\zeta}\sim 1<\frac{1}{\epsilon_V^2}$, to be treated perturbatively. Interestingly, this is not the case when the scalar field has a small sound speed. When $1-c_s^2>\epsilon$ the restriction from perturbativity (now using $\partial\rightarrow aH/c_s$) is instead \cite{Leblond:2008gg}
\begin{equation}
\mathcal{P}_{\zeta}<c_s^4<1\,.
\label{eq:perturbcs}
\end{equation}
The restriction is stronger since the dominant terms in the cubic action for the perturbations are inherited from the higher derivative terms in the scalar field action, and scale as $1/c_s^2$. Although from a low-energy effective field theory point of view, one expects $c_s\sim 1$ \cite{Creminelli:2003iq}, a well-known example with $c_s\ll1$ is DBI inflation \cite{Silverstein:2003hf}\footnote{Arguments based on an effective theory of the fluctuations alone allow small $c_s$ to appear generic \cite{Cheung:2007st}.}. Since only inflation models with $c_s\approx1$ have eternally inflating regimes that can be treated perturbatively, we restrict the discussion of whether eternally inflating models can be consistent with the swampland criteria to slow-roll scenarios.

\section{Eternal inflation and the refined de Sitter conjecture}

In the slow-roll case with $c_s=1$, the potentials that satisfy the conditions of the refined dS conjecture do not have slow-roll parameters that all scale as powers of $\epsilon_V$ and so we must revisit the requirements for stochastic eternal inflation to be under perturbative control. The second derivative of the potential is typically encoded in the second slow-roll parameter
\begin{equation}
\eta_V=M_p^2\left(\frac{V^{\prime\prime}}{V}\right)\,.
\end{equation}
Then, from Eqs.~(\ref{eq;conjecture1}) and (\ref{eq;conjecture2}), the potentials of interest for the dS conjecture may have $|\eta|\gg\epsilon$, in which case terms of order $\eta_V$ may dominate the cubic action and dictate the perturbativity condition.

To determine the conditions for a perturbative calculation of the power spectrum, we return to the full cubic action for the perturbations. For single-field inflation, including the effects of derivative self-interactions of the field, the expression can be found a convenient form in \cite{Adshead:2011bw}, which rewrites the results from \cite{Chen:2006nt}, which in turn builds on the calculation for $c_s=1$ from \cite{Maldacena:2002vr}. Since small sound speed in general only increases the size of higher order terms in the action, we consider the $c_s=1$ case:
\begin{align}
S_3=\int \d t \d^3 x &\left[a^3\epsilon_H\zeta\dot{\zeta}^2+a\epsilon_H\zeta(\partial\zeta)^2- 2a\epsilon_H\dot{\zeta}\partial_i\zeta\partial_i\chi\right.\nonumber\\
&+a^3\epsilon_H(\dot{\epsilon}_H-\dot{\eta}_H)\mathcal{\zeta}^2\dot{\mathcal{\zeta}}+ \frac{\epsilon^2_H}{2}a\partial_i\zeta\partial_i\chi\nonumber\\
&\left.-\frac{\d}{\d t}\left(a^3\epsilon_H(\epsilon_H-\eta_H)\mathcal{\zeta}^2\dot{\mathcal{\zeta}}\right)\right]\,
\end{align}
where $\chi=a^2\epsilon_H\partial^{-2}\dot{\zeta}$. The slow-roll parameters used here are
\begin{align}
\epsilon_H&=2M_p^2\left(\frac{H^{\prime}}{H}\right)^2\nonumber\\
\eta_H&=2M_p^2\frac{H^{\prime\prime}}{H}=\epsilon_H-\frac{1}{2H}\frac{\dot\epsilon_H}{\epsilon_H}\,.
\end{align}
Notice that the $\dot\eta$ term in the second line of $S_3$ cancels the similar term in third line. Since $\dot{\epsilon}_H$ can be written in terms of $\epsilon_H$ and $\eta_H$, we can estimate the size of the cubic action without the need for higher order terms in the slow-roll expansion. If higher order terms are unusually large, they may result in perturbativity conditions from higher order terms in the action imposing stronger constraints than those from $S_3$.

Ignoring the non-local terms that contain $\chi$ (whose inclusion will not qualitatively change our conclusions), the perturbativity condition estimated from $S_3/S_2<1$ gives
\begin{align}
& 2\mathcal{P}_{\zeta}^{1/2}&\left[-4\epsilon_H+2\epsilon_H^2-4\epsilon_H\eta_H+6\left(\eta_H+\frac{\eta_H^2}{3}\right)\right] < 1\nonumber\\
&\Rightarrow  \mathcal{P}_{\zeta}&<\frac{1}{16\left[-2\epsilon_H+\epsilon_H^2-2\epsilon_H\eta_H+3\left(\eta_H+\frac{\eta_H^2}{3}\right)\right]^2}
\end{align}
To put this in terms of derivatives of the potential, use
\begin{align}
\epsilon_V&=\epsilon_H\left(\frac{3-\eta_H}{3-\epsilon_H}\right)^2\approx\epsilon_H\\\nonumber
\eta_V&=\frac{1}{3-\epsilon_H}(3\epsilon_H+3\eta_H +\eta_H^2-\xi_H)\approx \epsilon_H+\eta_H
\end{align}
where $\xi_H=4M_p^4\frac{H^{\prime}H^{\prime\prime\prime}}{H^2}$ is the next order slow-roll parameter, and the final approximations hold for $\epsilon_H\ll1$, $|\eta_H|\ll3$, $\xi_H\ll\epsilon_H$,$\eta_H$.
Then, the perturbativity requirement can be written
\begin{align}
\mathcal{P}_{\zeta}&<\frac{1}{16[3(\eta_V-\epsilon_V)-\epsilon_V(\eta_V-\epsilon_V)-\epsilon_V\eta_V]^2}\nonumber\\
\mathcal{P}_{\zeta}&<\frac{1}{16\cdot9\eta_V^2}\;,\;\;\;\; {\rm for}\;|\eta_V|\gg\epsilon_V.
\label{eq:perturbeta}
\end{align}
Clearly, even allowing fairly generous numerical wiggle room for the approximation, $\mathcal{P}_{\zeta}\sim1$ is not compatible with the restrictions on $\epsilon_V$ and $\eta_V$ from the refined dS conjecture which requires $|\eta_V|\gtrsim\mathcal{O}(1)$. That is, potentials where $|\eta_V|\gg\epsilon_V$ that can be treated perturbatively in the stochastic regime are not sufficiently tachyonic to satisfy the criteria in Eq.(\ref{eq;conjecture2}), unless the criteria could be relaxed to allow $|\tilde{c}|<\mathcal{O}(\frac{1}{10})$.

It is worth discussing in a bit more detail the calculations for eternal inflation in the context of the class of potentials which are compatible with the refined dS conjecture. Following the analysis in \cite{Kinney:2018kew}, we consider potentials that satisfy the condition in Eq.~(\ref{eq;conjecture2}). These potentials support inflation at a maxima or inflection point of the potential. Expanded about the point where $V^{\prime}(\phi_0\equiv0)\equiv \frac{\d V}{\d\phi}\left.\right|_{\phi=\phi_0\equiv0}=0$ (and possibly $V^{\prime\prime}(\phi_0)=0$) the potential can be written as
\begin{equation}
V(\phi)=V_0\left[1-\lambda_p\left(\frac{\phi}{\mu}\right)^p+\dots \right]
\label{eq:hilltop}
\end{equation} 
for $p > 1$ and $\mu$ a parameter with units of mass, and we have assumed $\phi/\mu\ll1$. Inflection point potentials arise if two terms in the expansion in $\phi/\mu$ have coefficients that allow them to cancel at $\phi=0$. Additional discussion of these potentials in the context of the swampland and outside of the eternal regime can be found in \cite{Kinney:2018nny, Lin:2018rnx}.

When considering eternal inflation in these potentials with a tachyonic direction, there is an additional requirement, beyond Eq.(\ref{eq:dQdcl}), since quantum fluctuations that move the field by a distance comparable to the width $\Delta$ of the inflating plateau in field space will end inflation (hilltop potential), or are as likely to end inflation as to prolong it (inflection point potential). Given the potential, one can compute the field range $\Delta$ that supports (possibly eternal) inflation. The average time for inflation to end is when the typical distance the fluctuations have moved the field along the potential is of order $\Delta$. If the typical field motion in one Hubble time is $\langle\delta\phi\rangle_{\rm Q}$, then the typical time for inflation to end is $\langle t\rangle_{\rm end}\approx \Delta/(H \langle \delta\phi\rangle_{\rm Q})$. Demanding that the probability to generate new inflating Hubble volumes outweighs the probability to end inflation as fluctuations drive the field off the plateau gives an additional constraint on the parameters of the potential, which is approximately \cite{Barenboim:2016mmw}
\begin{equation}
\frac{\Delta}{\langle\delta\phi\rangle_Q}>\frac{1}{\sqrt{3}}.
\label{eq:plateau}
\end{equation}

To determine when stochastic eternal inflation can occur on a hilltop or inflection point potential, we must be able to compute $\langle\delta\phi\rangle_{\rm Q}$ to use in Eq.(\ref{eq:plateau}). The standard calculation is to add up the mean contributions of all momentum-mode fluctuations that cross out of the Hubble volume in one Hubble time (see, e.g, \cite{Linde:2005ht}). If the fluctuating field is well-approximated by a massless, free field, the power in fluctuations clearly comes only from the quadratic action, and performing the integral gives the standard result $\langle\delta\phi\rangle_Q=\frac{H}{2\pi}$. If this result for $\langle\delta\phi\rangle_Q$ is valid during stochastic eternal inflation on a plateau potential with, for example $p=2$, Eq.(\ref{eq:plateau}) restricts the the parameters in the potential to satisfy \cite{Barenboim:2016mmw}
\begin{align}
{\rm Assuming\;\,perturbativity}:\;\;\;\;\left(\frac{\mu}{M_p}\right)^2 >  \frac{2}{\sqrt{3}}
\label{eq:p2}
\end{align}
which implies $\eta>-\sqrt{3}$. Since Eq.(\ref{eq;conjecture2}) says $\eta<-c$ for $c\sim\mathcal{O}(1)$, \cite{Kinney:2018kew} used Eq.(\ref{eq:p2}) to conclude that stochastic eternal inflation is marginally consistent with the refined dS conjecture. For larger values of $p$ in hilltop potentials, as long as we include the $p = 2$ term, \cite{Barenboim:2016mmw} showed that $\eta$ is in general bounded by a larger negative number. If we keep only $p>2$ terms, then Eq.(\ref{eq;conjecture2}) is automatically violated (while  Eq.(\ref{eq;conjecture1}) remains valid) near the hilltop, and one gets a potential which is inconsistent with the refined dS conjecture.

However, when the perturbativity constraint is violated, the field fluctuations are far from Gaussian, one cannot calculate $\langle\delta\phi\rangle_{\rm Q}$ using the quadratic result for the power spectrum, and so Eq.(\ref{eq:p2}) can no longer be trusted as the criteria for stochastic eternal inflation on a hilltop potential \footnote{Note that here the usual Fokker-Planck, or Langevin, equation familiar from \cite{Starobinsky:1986fx,Guth:1985ya} may not be used. Those equations assume that a truncation of the master equation at two derivatives is valid and are not a consistent starting point if the non-Gaussian corrections to the two-point function are already large in one Hubble volume and one Hubble time.}. In this case, we fall back on the necessary, but not sufficient, requirement for stochastic eternal inflation, $\mathcal{P}_{\zeta}\sim1$, to demonstrate that perturbative control of the stochastic regime is incompatible with $|\eta_V|\gtrsim\mathcal{O}(1)$ and the refined dS conjecture.

\section{Perturbative, stochastic eternal inflation is in the swampland}

The argument of \cite{Ooguri:2018wrx} can be made including the possibility of a stochastic eternal inflation regime, as long as we have a good semi-classical description and can define an event horizon that contains the apparent horizon at all times. The difference from the original construction is that allowing for a stochastic regime means that quantum fluctuations prevent the field in a single observer's volume from following the classical trajectory for at least some period of time, and so the position of the apparent horizon deviates from that in the slow-roll regime. However, it has been argued in \cite{Bousso:2006ge} that there is still a controlled local description in volumes that undergo some eternal inflation, but where inflation eventually ends. Assuming the local amplitude of quantum fluctuations is calculable, which requires the perturbativity conditions derived above, \cite{Bousso:2006ge} showed that the change in the area of the apparent horizon due to the quantum-fluctuation dominated regime is small. The event horizon set by the value of $H$ at the end of inflation in the volume will then still provide a good volume to use to apply the covariant entropy bound. In cases where this argument does not apply, stochastic eternal inflation is even further beyond calculational control than what is indicated by the perturbativity conditions.

So, we may begin with the assumption of a quasi-dS volume that is allowed to explore regions of the potential where $\mathcal{P}_{\zeta}>1$. The criteria to apply the covariant entropy bound to this scenario, based on a semi-classical event horizon defined to be at all times outside the apparent horizon are then not only the Ooguri et al criteria $V^{\prime\prime}\geq -\tilde{c}\, V$ but also that the perturbativity constraints, including (but not limited to) the strongest of  Eq.(\ref{eq:perturbeps}), Eq.(\ref{eq:perturbcs}), Eq.(\ref{eq:perturbeta}). If these constraints are violated, we cannot calculate where the apparent horizon lies when the quantum fluctuations drive the evolution of the field sourcing inflation. 

The consistency of the distance conjecture \footnote{There may be additional contributions to the entropy from the modes that are red-shifted into the low-energy theory during the stochastic regime, but this can only act to further tighten the bound. It would be interesting to understand how relevant such a contribution might be.} and covariant entropy bound, applied to quasi-dS spacetimes sourced by a single-field, which experiences inevitable quantum fluctuations, constrains both the potential and the behavior of quantum fluctuations of the field. That is, when
\begin{equation}
|V^{\prime}|<\frac{c}{M_p}V\nonumber
\end{equation}
then the low energy effective theory descending from a consistent quantum gravity theory must satisfy constraints on both the potential and the fluctuations:
\begin{align}
V^{\prime\prime}&\leq\frac{-\tilde{c}}{M_p^2}V\nonumber\\
\;\;\;&{\rm\bf and}\;\;\;\nonumber\\
S^{(\zeta)}_n/S^{(\zeta)}_2&<1\;\;{\rm for\;\,all}\;\,n>2.
\label{eq:conjecture3}
\end{align}
Some version of the last condition (perturbative control) is, of course, often required in formulating swampland criteria. In the specific case of quasi-dS solutions, where it is naively possible to have significant semi-classical back-reaction from the quantum fluctuations sourced by the expansion, we have demonstrated an example of the misleading conclusions that can be drawn if the perturbativity constraint is not explicitly enforced. It would be interesting to extend this analysis to scenarios where additional degrees of freedom are relevant for the scalar dynamics, including the case of warm inflation \cite{Motaharfar:2018zyb,Das:2018rpg}, and to see if there are implications for scenarios where inflation is not driven by a fundamental scalar field \cite{Heckman:2019dsj}.

\section{Conclusions}

We have shown that single-field stochastic eternal inflation is beyond the regime of control in low-energy theories that descend from UV complete theories of quantum gravity, if the refined dS conjecture holds. (A similar result, from a different perspective, is advocated in \cite{Dvali:2018jhn}.) Following the reasoning of \cite{Ooguri:2018wrx}, the validity of our conclusion requires that two conjectures about the nature of quantum gravity are correct -- the distance conjecture and the covariant entropy bound -- as well as a controlled description of quantum fluctuations. Effects coming from the coupling of the inflaton to heavy degrees of freedom (which generically result in $c_s<1$) can only make the conclusion stronger. Stochastic eternal inflation has often been invoked as a means of dynamically generating a universe (or, multiverse) where different parts of the string landscape are populated \cite{Linde:2006nw,Clifton:2007en}. Most often, this is done to address the cosmological constant problem. We have shown that if the refined dS conjecture holds, especially beyond the perturbative regime, it not only restricts the landscape but also the extent to which quantum effects can be invoked to dynamically populate the landscape within a single universe.

\vspace{5cm}

\section*{Acknowledgements}
We thank Gary Shiu and Will Kinney for very helpful discussions. S.S. is supported by the National Science Foundation under award PHY-1719991. S.B. is supported in part by the Ministry of Science, ICT \& Future Planning, Gyeongsangbuk-do and Pohang City and the National Research Foundation of Korea (Grant No.: 2018R1D1A1B07049126).

\bibliographystyle{unsrt}
\bibliography{SwampEI}

\end{document}